\newcommand{\neff}{n_{e}}
\newcommand{\Dneff}{\Delta n_{e}}

\newcommand{\na}{n_{h}}
\newcommand{\Dna}{\Delta n_{h}}

\newcommand{\ns}{n_{s}}

\newcommand{\Domega}{\Delta\omega}

\newcommand{\vecR}{\hat{s}}

\newcommand{\vf}{\phi}
\newcommand{\vecr}{\mathrm{\mathbf{r}}}

\newcommand{\tn}{\tau_{\delta}}

\newcommand{\ls}{\ell_s}

\documentclass[aps,prl,twocolumn,showpacs]{revtex4}
\usepackage{amsmath,array,graphicx}
\begin{document}

\title{Varying the effective
refractive index to measure optical transport
in random media}
\author{Sanli Faez}
\email{faez@amolf.nl}
\affiliation {FOM Institute for Atomic and Molecular Physics AMOLF, Science Park 113,
1098 XG Amsterdam, The Netherlands}
\author{P. M. Johnson}
\affiliation{FOM Institute for Atomic and Molecular Physics AMOLF, Science Park 113,
1098 XG Amsterdam, The Netherlands}
\author{Ad Lagendijk}
\affiliation{FOM Institute for Atomic and Molecular Physics AMOLF, Science Park 113,
1098 XG Amsterdam, The Netherlands}

\begin{abstract}
We introduce a new approach for
measuring both the effective medium and the
transport properties of light
propagation in heterogeneous media.
Our method utilizes the conceptual equivalence of
frequency variation with a change in
the effective index of refraction. Experimentally, we
measure intensity correlations via spectrally resolved
refractive index
tuning, controlling the latter
via changes in the ambient pressure.
Our experimental results
perfectly match a generalized transport theory that incorporates
the effective medium and predicts a precise value for
the diffusion constant. Thus, we directly confirm the
applicability of the effective medium
concept in strongly scattering materials.
\end{abstract}



\maketitle

Light propagation in multiple-scattering media
is dominated by speckle, a
highly irregular intensity pattern
dependent upon spatial (or angular) or
time (or frequency) coordinates and
brought about by interference.
Correlations, which are inherent
properties of speckle despite the
apparent irregularity, provide
important information about transport
parameters. In fact, the description of
intensity correlations is at the heart
of understanding transport theory
\cite{feng_correlations_1988}. In the
past, correlations have been measured
in time~\cite{cai_timeresolved_1996}
and
frequency~\cite{genack_optical_1987}
as a means of, for example, determining
the diffusion constant of light.

What is often not fully appreciated is the degree
to which effective medium properties
are essential for determining
correlation functions, and thereby
transport properties. When developing a
theory of transport, the effective
medium is an essential building block~\cite{sheng_introduction_1995}.
In practice, fundamental problems arise when one
attempts to access the effective medium properties
of strongly scattering materials. In a turbid medium,
the coherent
propagation, upon which the effective medium
properties are defined, decays on the scale
of a few mean free paths and becomes unmeasurable in the overwhelming
bath of diffuse intensity. On the other hand, the
assumptions needed for developing standard effective
medium models do not hold for
such materials, which makes the models unapplicable.
These difficulties raise the question of
what exact role the effective medium
plays in determining transport
properties in strongly scattering
materials.

In this letter we first highlight the central role of the
effective refractive index in transport theory.
We then propose and measure a new
correlation function that directly incorporates the
effective medium approach. We validate this approach
by showing the precise agreement
between the theory and experimental results.
This comparison requires defining a new \emph{directly measurable} effective
medium parameter, which
we call the ``tuning response'', as
\begin{equation}\label{eq:definitiondelta}
\delta \equiv
\frac{1}{\neff}\frac{\partial
\neff}{\partial \na}.
\end{equation}
The parameter $\delta$ relates the
change of the effective refractive index $\neff$ to the changes in
the refractive index $\na$ of one of the components, which can be easily
experimentally controlled.  We measure the precise value of $\delta$
with our method and use it to extract the diffusion
constant. Finally, we use $\delta$ to test several standard models of the
effective medium.

Our measurement method is based on controlled
change of the optical path length
distribution by refractive index tuning (RIT).
It is the
optical analogue of a class of experiments in
condensed matter physics that control the elongation of electron
trajectories by using magnetic
field. Those experiments resulted in
the observation of electronic weak localization
and universal conductance
fluctuations~\cite{bergmann_1982}.
The concept of changing the path length
distribution has also been exploited for diffusing
wave spectroscopy~\cite{maret_multiple_1987,pine_diffusing_1988},
and for other types of waves in evolving media~\cite{lobkisweaver_2003, larose_2006,
page_phystoday_2007}. Given the predominance of disordered photonic media in a wide
variety of fields including
biology~\cite{gibson_2005}, advanced
materials~\cite{smith_science2004},
solar cells~\cite{muskens_design_2008},
and in general modern photonics, we
expect our approach to have broad
cross-disciplinary application.

In standard transport theory~\cite{sheng_introduction_1995}, the effective
refractive index shows up in the early stages where the the averaged amplitude
Green's function is introduced:
\begin{equation}\label{eq:average_amplitude_greens_function}
G_\omega(\vecr;\neff,\ls)=-\frac{
e^{\left( \frac{i \neff\omega}{c}
- \frac{ 1}{2 \ls}\right) r} } {4\pi
r},
\end{equation}
where $\ls$ is the scattering mean free
path. This form of the Green's function takes care
of internal resonances of scatterers but does not hold
in a regime of strong
spatial dispersion, for example in photonic crystals, or
when the material is anisotropic. One can introduce the effective
(complex-valued) wavenumber by
$ K\equiv \frac{2 \pi}{\lambda_e} +
i \frac{1}{2 \ls}.$

We introduce a new correlation function, which is central to our
measurements, defined as:
\begin{widetext}
\begin{eqnarray}
 C_{\omega,\omega + \Domega}(\neff, \neff +
\Dneff)\equiv
    N\left[\langle
I_\omega(\vecR;\neff)I_{\omega +
\Domega}(\vecR;\neff+\Dneff)\rangle
- \langle
I_\omega(\vecR;\neff)\rangle\langle
I_{\omega +
\Domega}(\vecR;\neff+\Dneff)\rangle
\right], \label{eq:c1}
\end{eqnarray}
\end{widetext}

where $I_\omega(\vecR)$ is the
far-field specific intensity at
direction $\vecR$ and $\langle.\rangle$
denotes averaging over a narrow spread
of directions. The normalization
constant $N$ is fixed by requiring
 $C_{\omega,\omega }(\neff, \neff) = 1$.
Intensity correlations
are typically classified into different
contributions: $C^{(1)}$, $C^{(2)}$, and $C^{(3)}$~\cite{feng_correlations_1988}.
Their relative
magnitude depends on the scattering strength of the material. We will present
the closed form for the first order contribution
$C^{(1)}$, which is sufficient to describe our
measurements,
but our argument holds for the higher-order terms as well.

The calculation of correlation
function~(\ref{eq:c1}) for the case of
$\Dneff = 0$ can
be found in many papers and
textbooks~\cite{sheng_introduction_1995,RMP99}.
The actual expressions depend
on the geometry. Closed forms have been
presented for the case of an infinite
medium~\cite{Shapiro_PRL86},
semi-infinite
medium~\cite{freund_1990}, and a
slab~\cite{genack_optical_1987}. Yet
conventional definitions of this type
of correlation function have not
included variations of $\neff$. The
crucial point of the theoretical part
of the present paper is that the
formulae calculated for frequency
correlations can easily be generalized
to the case of $\Dneff \ne 0$.

In our regime of consideration where
\begin{equation}\label{eq:changeimk}
\frac{\Delta \mathrm{Im}K}{\Delta
\mathrm{Re}K} \sim \frac{1}{4
\pi}\left(\frac{\lambda_e}{\ls}\right)^2\left|\frac
{\Delta \ls}{\Delta \lambda_e}\right| \ll
1,
\end{equation}
we can neglect variations of $\rm{Im}K$
in the average Green's functions. As a
result, the change in the average Green's
function~(\ref{eq:average_amplitude_greens_function})
is only dependent on the product $n_e \omega $.
That is to say changing the frequency
by $\Delta \omega $ is equivalent to
changing $\neff$ by $\Dneff$ if $\neff
\Domega = \omega \Dneff $.

\begin{figure}[b]
  \includegraphics[width=6cm]{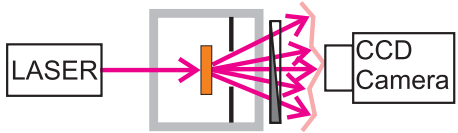}
  \caption{The experimental setup. The sample is placed in pressurized chamber
  and illuminated by a HeNe laser. A single polarization of the
  transmitted speckle is recorded by a CCD-camera.}
  \label{fig:setup2}
\end{figure}

We prove that the same symmetry between
frequency and effective refractive index variation
holds for the intensity correlation
function~(\ref{eq:c1}). The basic ingredients for
deriving this correlation function are
the average Green's function~(\ref{eq:average_amplitude_greens_function}) and an
irreducible scattering
vertex~\cite{sheng_introduction_1995}.
We symbolically denote the irreducible
scattering vertex as: $U(\omega,\omega
+ \Domega;\neff, \neff + \Dneff)$,
and assume that it is local, thus we have dropped the three
momentum variables. Changes caused by
$U$ are compensated by
those resulting from $\Delta \mathrm{Im}K$ due to energy
conservation. This cancelation holds even
without considering
condition~(\ref{eq:changeimk}).
As a result, replacing
$\neff (\omega + \Delta \omega)$ by $(\neff +
\Dneff )( \omega + \Domega)$ does not affect
any algebraic step
in the derivation of $C_{\omega,\omega +
\Domega}(\neff, \neff)$~\cite{RMP99} and thus yields the
generalized correlation function
$C_{\omega,\omega + \Delta
\omega}(\neff, \neff +\Dneff)$.

\begin{figure}[t]
  \includegraphics[width=6cm]{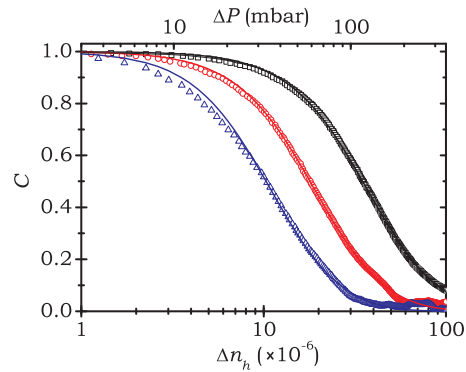}
  \caption{\label{fig:c1correlation}RIT autocorrelation coefficient as a function of $\Dna$. Data points
  show measurement results for 3 slabs of polyethylene filters ($L=1,1.5,$ and 2~mm for
  black squares, red circles, and blue triangles, consequently).
  The lines show the fit to theory (Eq.~(\ref{eq:c1slab})).}
\end{figure}

To take one example consider a slab of porous material,
for which
the host refractive index, $\na$, may be tuned.
The correlation is then measured as a
function of $\na$ instead of $\neff$. In
this case, $C^{(1)}_{\omega,\omega } (
\na, \na + \Dna)$ is given by:
\begin{eqnarray}
\label{eq:c1slab}
C^{(1)}_{\omega,\omega } ( \na, \na +
\Dna)=
\frac{\tn \Dna}{\cosh{\sqrt{\tn\Dna}}-\cos{\sqrt{\tn\Dna}}},
\end{eqnarray}
with RIT decay coefficient (generalized analogue of diffuse decay time)
\begin{eqnarray}
\tn\equiv\frac{2
\omega \delta L^2}{D},
\end{eqnarray}
where we have used the tuning
response $\delta$ defined in
Eq.~(\ref{eq:definitiondelta}).

\begin{figure}[t]
  \includegraphics[width=6cm]{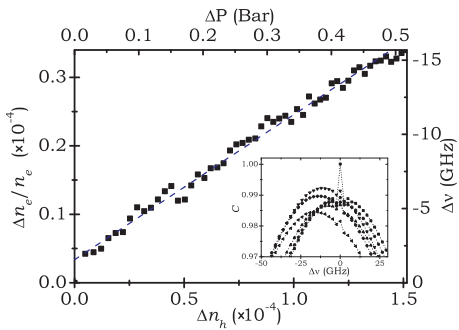}\\
  \caption{\label{fig:memory}
  Measured relative change of $\neff$ as a function of $\Dna$.
  The value of $\Dneff/\neff$ is extracted by measuring this
  spectral shift. The refractive index of air is calculated based on the Edlen formula~\cite{edlen_refractive_1966}.
   Inset: The shifted cross-correlation coefficient $C$ versus the spectral shift
  $\Delta \nu$ for certain pressures (From right to left:
  $\Delta P=$ 0.0, 0.075, 0.18, 0.28, 0.37, and 0.46 bar).  The peak position of each curve
  denotes the spectral shift, which is proportional to the relative change in the effective
  refractive index. The spiky feature at zero position is an artifact of the
  slight inhomogeneity over the detection efficiency of CCD pixels.
    }
\end{figure}

We apply our method to typical samples, which are representative for a
a large class of multiple scattering
materials. The sample dimensions are selected such
that they can be characterized by
both RIT and time-resolved
measurements, allowing us to show the
accuracy of our technique.
We have used a composite
material with open channels consisting of
a solid backbone and a gaseous host.
The refractive indices of the solid
backbone and the gaseous host medium
are $\ns$ and $\na$ respectively. In
our measurements the index of the host is tuned by increasing the
pressure of the gas.

As samples we used slabs of
commercially available porous plastic
air-filters (XS-7744, Porex Corp.)
available in different thicknesses.
They are composed of sintered
polyethelene spheres with a broad size
distribution of 7 to 12~$\mu$m. The
refractive index of the polyethelene is
$\ns=1.49$. This material has a porosity of
$\vf=0.46\pm0.02$ and a
mean free path of
$\ls=20.6\pm0.2\;\mu$m in the frequency range of interest. The
porosity was determined by weighing a
larger sample of the same material.
The mean
free path was determined from total
transmission measurements. These
measurement also revealed a weak
dependence of $\ls$ on frequency,
justifying the assumption~(\ref{eq:changeimk}).

The sample was kept in a pressurized
chamber and illuminated by a HeNe laser
at 632.8~nm. Part of the transmitted
speckle pattern was filtered by a
linear polarizer and recorded on a
16-bit
CCD-camera with $10^6$ pixels. The
recorded image consisted of $\sim10^4$
independent coherence areas.
Gradually tuning the air pressure in
the chamber changes $\na$, causing the
speckle pattern to evolve.
This evolution allowed
us to directly measure the
autocorrelation coefficient defined in
Eq.~(\ref{eq:c1}).

This experiment was performed on three
samples. The extracted autocorrelation
coefficients are plotted in
Fig.~\ref{fig:c1correlation} versus
$\Dna$. Each data series is fitted to
the correlation function of
Eq.~(\ref{eq:c1slab}) with a separate
single fitting parameter, $\tn$. We see
an excellent agreement between the
theory and the experimental results. This agreement is an important
confirmation of our theory which is built upon
the concept of effective wavenumber in random media. This concept
has never been examined so directly in strongly scattering media, simply because
it was a very difficult parameter to access experimentally.

Our experimental results not only match the
theoretically predicted functionality, but are
quantitatively precise. We show this
precision by extracting the diffusion
constant from the RIT decay
coefficient $\tn$ and comparing it with
the diffusion constant measured by an standard
method. This requires the
value of $\delta$ which we can measure
using the symmetry relation discussed earlier in
the paper. This symmetry implies that
$C_{\omega,\omega+\Domega}(\neff,\neff+\Dneff)$
has a peak equal to one for a nonzero
shift $\Domega$ given by
\begin{equation}\label{eq:lambdamemory}
    ({\neff+\Dneff})({\omega+\Domega})={\neff}{\omega}\;\;\;\textrm{or}\;\;\;
    \frac{\Domega}{\omega}=-\frac{\Dneff}{\neff}.
\end{equation}
In other words, $\delta$ can be
extracted from the spectral
shift of every individual speckle spot
while the host index of refraction
$\na$ is varied.

To monitor this spectral shift the
experimental setup was changed. The
light source was replaced by a
white-light super-continuum laser
(Fianium). The CCD-camera was replaced
by a spectrometer, which was run in the
imaging mode. The entrance slit of the
spectrometer selects a transmission
direction in form of a narrow
rectangle, which contains roughly 10
independent coherence areas along the
slit. The beam is spectrally resolved
perpendicular to the slit direction by
a grating. This configuration allows us
to simultaneously monitor the spectral
evolution of several speckle spots
while $\na$ is changed.

The measurement was performed on a 100-$\mu$m-thick
slice of the polyethelene filter. A
thin sample was necessary due to the
limited
resolution of our spectrometer.
The intensity spectrum between 649 and
651~nm was measured while changing the
pressure in the chamber from 1 to 1.5
bar in steps of 10~mbar.

RIT correlation functions at different frequencies
are calculated
from the measurements and plotted as a
function of $\Domega$ for each
pressure. A collection of these plots
is shown for six different pressures
in the inset of Fig.~\ref{fig:memory}
as typical representatives. As
predicted, the peak value is shifted
and equals unity within the
experimental error. Using
our symmetry relation~(\ref{eq:lambdamemory}), the
relative change of effective refractive
index is calculated and plotted as a
function of $\Dna$ in
Fig.~\ref{fig:memory}. We get a
value of $\delta=0.212\pm0.003$ from the slope
of a linear fit to the data, for these set of samples,
with a remarkably high precision.

\begin{figure}[t]
  \includegraphics[width=6.0cm]{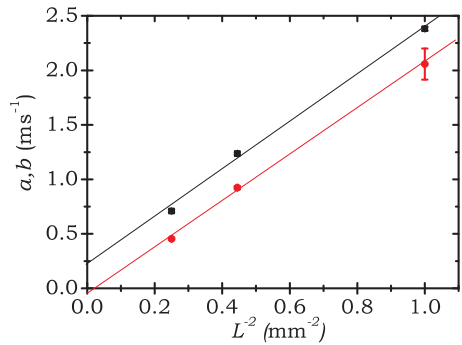}\\
  \caption{\label{fig:line}The measurement parameters $a\equiv\frac{4 \pi \nu \delta}{\tn}$
  and $b\equiv\frac{1}{\pi^2\tau_d}$ are plotted versus
  the inverse of the thickness squared. Black squares and red dots correspond to the
  index tuning and time-resolved measurements respectively.
  The slope of each data set is equal to the diffusion constant $D$
  measured by each specific method.  The fact that the two slopes
  are equal proves the consistency of our index tuning method
  with the time resolved measurements.}
\end{figure}

In principle, it is possible to extract the diffusion constant from
each separate RIT measurement.
To improve the accuracy, we plot
the auxiliary parameter $a\equiv{4 \pi
\nu \delta}/{\tn}$ versus $L^{-2}$ from
the three measurements presented in Fig.~\ref{fig:c1correlation}.
This plot is shown in
Fig.~\ref{fig:line}. The slope gives
the diffusion constant $D=2.2\pm0.1\;
\textrm{m}^2/\textrm{ms}$ using
Eq.~(\ref{eq:c1slab}).

Next we compare our result with the
diffusion constant measured by conventional time-resolved
method. The diffuse transmission
through the same samples was recorded using
a time-correlated photon counting setup
and a subpicosecond pulsed laser at
600~nm wavelength. The resulting temporal decay
curves are presented in the
supplementary information. Following
the guidelines of~\cite{johnson_2003}
and considering the response function
of the detector, the diffuse decay time
$\tau_d\equiv L^2/\pi^2 D$ is extracted
from a fit to these measurements. These
results confirm that the
absorption has a negligible effect on our experiments.
The diffusion constant is calculated
from the slope of a linear fit of the
parameter $
b\equiv\frac{1}{\pi^2\tau_d}.$ vs.
$L^{-2}$ (Fig.~\ref{fig:line}),
$D=2.1\pm0.1\;\textrm{m}^2/\textrm{ms}$,
which is in excellent agreement with the
result from the index tuning
experiment.

Our precise measurement method for measuring
$\delta$ is applicable to a variety of
samples, and assumes
no specific effective medium model.
The parameter $\delta$ can
also be predicted using models of
the effective medium~\cite{bohren_book,sheng_scalar_1986,busch_transport_1995}
which are all based on
long-wavelength limits.

For illustrative purposes, we present the
predicted value for $\neff$ and $\delta$ by three of the more
popular effective medium
theories: (i) the average
permittivity~\footnote{$\neff^2=\vf\na^2+(1-\vf)\ns^2$},
(ii) the Maxwell-Garnett
and (iii) the
Bruggeman models.
The predictions are
$\neff=1.29\pm0.02,\,1.25\pm0.02,$ and
$1.22\pm0.02$ and
$\delta=0.28\pm0.02,\,0.45\pm0.03,$ and
$0.52\pm0.03$ respectively. These
predictions for $\delta$ differ from
each other and from
our experimental result by as much as a
factor of two. It is also worth noting
that the predicted value of $\delta$ is
more model-dependent than
$\neff$ itself. The main reason for this discrepancy
is perhaps the assumption of the long
wavelength limit, which is not reached
in our experiments. Our unambiguous measurement of $\delta$
highlights the necessity of more sophisticated
models of the effective
medium~\cite{busch_transport_1995}.

In conclusion, we have presented a new effective medium quantity and a new transport
correlation function and shown how to precisely measure both of them by using refractive
index tuning. Our measurements directly test and approve the validity of
assuming an effective wavenumber for an inhomogeneous medium, which is very important for
describing all sorts of photonic metamaterials.
Using these two quantities one can measure several dynamic
transport properties with high precision, as we have demonstrated for
the lowest order $C^{(1)}$
correlations and the diffusion constant.
Measuring higher order correlations and
studying Anderson localization is a
natural follow-up to our research.

We thank D. Spaanderman for
designing the pressure chamber. This
work is part of the research program of
the ``Stichting voor Fundamenteel
Onderzoek der Materie'', which is
financially supported by the
``Nederlandse Organisatie voor
Wetenschappelijk Onderzoek''.

\newpage

\begin{appendix}


\begin{widetext}
\section{Supplementary information}
\begin{figure}[htb]
\includegraphics[width=8.3cm]{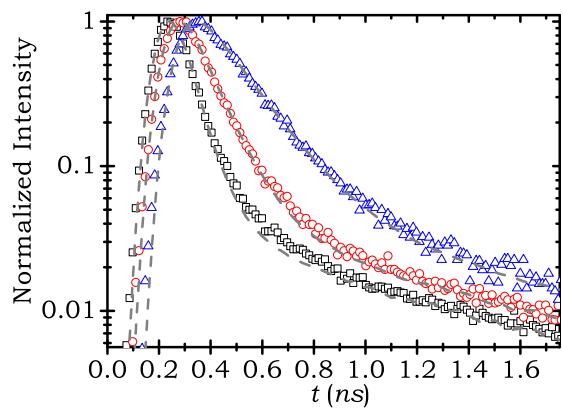}
  \caption{\label{fig:timeresolved}Time resolved intensity decay curve for slabs
  of polyethylene filters.
  Different datasets correspond to different
  sample thicknesses ($L=1,1.5,$ and 2~mm for
  black squares, red circles, and blue triangles, respectively).
  The lines show the fit to the diffusion model after considering the absorption and
  the response of the detector.}
\end{figure}
\end{widetext}
\end{appendix}

\end{document}